\documentclass[fleqn,useAMS,usenatbib]{mn2e}
\usepackage{epsfig,subfigure,amsmath,graphicx}
\usepackage{color}

\definecolor{purple}{rgb}{1,0,1}

\newcommand{\lcdm}{$\Lambda$CDM}
\newcommand{\hmpc}{$h^{-1}$Mpc}
\newcommand{\gmpc}{$h^{-1}$Gpc}
\newcommand{\hmsol}{\mbox{ } {h}^{-1}~{M}_{\odot}}

\DeclareMathOperator\erf{erf}

\bibpunct[; ]{(}{)}{;}{a}{}{,}

\title{Sparse sampling, galaxy bias, and voids}

\author[P.~M. Sutter et al.]
{
\parbox{\textwidth}{
{P.~M. Sutter}$^{1,2,3,4}$ \thanks{Email: sutter@iap.fr},
Guilhem Lavaux$^{1,2,5,6,7}$,
Nico Hamaus$^{1,2,4}$, 
Benjamin D. Wandelt$^{1,2,4,8}$,
David H. Weinberg$^{3,9}$, and
Michael S. Warren$^{10}$
}
\vspace{0.4cm}\\
\parbox[c]{\textwidth}{
$^{1}$ Sorbonne Universit\'{e}s, UPMC Univ Paris 06, UMR7095, Institut d'Astrophysique de Paris, F-75014, Paris, France \\
$^{2}$ CNRS, UMR7095, Institut d'Astrophysique de Paris, F-75014, Paris, France \\
$^{3}$ Center for Cosmology and Astro-Particle Physics, Ohio State University, Columbus, OH 43210\\
$^{4}$ Department of Physics, University of Illinois at Urbana-Champaign, Urbana, IL 61801\\
$^{5}$ Department of Physics \& Astronomy, University of Waterloo, Waterloo,
ON,  N2L 3G1 Canada \\
$^{6}$ Perimeter Institute for Theoretical Physics,
Waterloo, ON, N2L 2Y5, Canada \\
$^{7}$ Canadian Institute for Theoretical Astrophysics, 60 St. George St.,
Toronto, ON M5S 3H8 Canada \\
$^{8}$ Department of Astronomy, University of Illinois at Urbana-Champaign, Urbana, IL 61801\\
$^{9}$ Department of Astronomy, Ohio State University, Columbus, OH 43210\\
$^{10}$ Theoretical Division, Los Alamos National Laboratory, Los Alamos, NM 87545, USA
}}

\begin{document}

\maketitle

\label{firstpage}

\begin{abstract}
To study the impact of sparsity and galaxy bias on void statistics,
we use a single large-volume, high-resolution $N$-body simulation
to compare voids in multiple levels of subsampled dark matter, 
halo populations, and mock galaxies from a Halo Occupation Distribution 
model tuned to different galaxy survey densities. 
We focus our comparison on three key observational statistics:
number functions, ellipticity distributions, and radial density 
profiles.
We use the hierarchical tree structure of voids 
to interpret the 
impacts of sampling density and galaxy bias, and
theoretical and empirical functions to describe the 
statistics in all our sample populations.
We are able to make simple adjustments to theoretical 
expectations to offer 
prescriptions for translating from analytics to the void properties 
measured in realistic observations.
We find that sampling density has a much larger effect on void 
sizes than galaxy bias. At lower tracer density, 
small voids disappear and the remaining voids are larger, 
more spherical, and have slightly steeper profiles.
When a proper lower mass threshold is chosen, 
voids in halo distributions largely mimic those found in galaxy populations,
except for ellipticities, where galaxy bias leads to 
higher values.
We use the void density profile 
of Hamaus et al. (2014) to show that voids follow 
a self-similar and universal trend, allowing simple translations 
between voids studied in dark matter and voids 
identified in galaxy surveys.
We have added the mock void catalogs used in this work
to the Public Cosmic Void Catalog at http://www.cosmicvoids.net.
\end{abstract}


\section{Introduction}
Voids --- the large, underdense structures that fill up most of the 
volume of the universe --- are a unique and potentially powerful 
cosmological probe~\citep{Thompson2011}. 
Their size and shape distributions are sensitive to
the nature of dark energy~\citep[e.g.,][]{Biswas2010,Bos2012,Li2012, Shoji, Jennings2013}, 
their internal dynamics are strongly altered
by fifth forces and modified gravity~\citep[e.g.,][]{Li2009,Clampitt2013,Spolyar2013},
and the integrated Sachs-Wolfe effect offers constraints on 
cosmological parameters~\citep{Thompson1987,Ilic2013,Planck2013b}.
Their statistical isotropy means that we can use stacked voids 
as a standard sphere for an Alcock-Paczynski 
test~\citep{Ryden1995, LavauxGuilhem2011, Sutter2012b}.
Additionally, we may use voids to test for the existence of 
primordial magnetic fields~\citep{Taylor2011,Beck2013}
and study the effects of environment on 
galaxy evolution~\citep{Gottlober2003,VandeWeygaertR.2011,Ceccarelli2012}.

Researchers typically use analytical calculations and $N$-body dark matter 
simulations to predict and understand various 
void properties, such as 
number functions~\citep{Sheth2004, Furlanetto2006,Paranjape2012,Jennings2013, Achitouv2013}, 
radial density profiles~\citep{Fillmore1984,Dubinski1993,Benson2003,Colberg2005,Ceccarelli2006}, 
and ellipticities~\citep{Lavaux2010,Bos2012,Shoji}.
However, absent direct measurements of dark matter underdensities 
the only way to infer the characteristics of
 voids is with large galaxy redshift 
surveys~\citep{Pan2011, Sutter2012a}.
These galaxy populations are sparse, biased tracers of the underlying 
dark matter density field, which can potententially impact 
void statistics.

There has already been some work to make the connection between 
galaxy and dark matter voids:
~\citet{Furlanetto2006} and~\citet{Jennings2013} made 
parameter adjustments to the excursion set formalism of~\citet{Sheth2004}
to account for void finding in galaxy populations,
~\citet{Ryden1996} compared voids in real and redshift space,
~\citet{Tinker2009} looked at the impacts of the biasing of different 
galaxy populations,
and~\citet{Schmidt2001} and~\citet{Colberg2005} made early assessments of the 
effects of different sparsity levels on void reconstruction.
When attempting to \emph{predict} void statistical properties, most authors 
find voids within the halo distribution, not a mock galaxy population~\citep[e.g.,][]{Bos2012,Jennings2013}.
We will discuss the validity of this assumption below.
When attempting to \emph{match} observed void statistics, authors 
typically turn to semi-analytic modeling~(SAM;~\citealt{DeLucia2009}) to generate 
mock galaxy populations~\citep{Benson2003,Padilla2005,Ceccarelli2006,Pan2011,Tavasoli2013}.

While it may be possible to link galaxy voids to dark matter 
voids on a one-by-one basis, this is very difficult due to the 
complex internal hierarchical 
structure of voids~\citep{Aragon2012, Sutter2013b}.
Instead, in this work we take a holistic approach and offer 
general prescriptions to translate from dark matter voids to 
galaxy voids by examining voids in different tracer populations 
sourced from the same cosmological simulation. 
We focus on three key void observables: 
number functions, ellipticity distributions, and radial 
density profiles. We take theoretical- and empirically-derived 
fitting functions and adjust their parameters to move smoothly 
among the different populations.

We begin in Section~\ref{sec:approach} with our simulation setup, 
our process for generating mock galaxy populations, and 
our void finding algorithm. Next in Section~\ref{sec:tree} 
we use the hierarchical tree structure of voids to examine the 
consequences of reducing the tracer density and introducing 
galaxy bias. Section~\ref{sec:observable} discusses the 
changes to number functions, ellipticity distributions, 
and radial density profiles. Finally, we conclude in 
Section~\ref{sec:conclusions} with a discussion of consequences for 
interpreting voids in galaxy redshift surveys
and comments on future work.

\section{Numerical Approach}
\label{sec:approach}
\subsection{Simulations \& Mocks}
We source all samples and mock catalogs in this work from a single 
\lcdm~dark matter $N$-body simulation. 
We use the 
{\tt 2HOT} code, an adaptive treecode N-body method whose operation count
scales as $N \log N$ in the number of particles~\citep{warren13}.
Accuracy and error behavior have been improved significantly for
cosmological volumes through the use of a technique to subtract the
uniform background density, as well as using a compensating smoothing
kernel for small-scale force softening~\citep{dehnen01}.  We use a
standard symplectic integrator~\citep{quinn97} and an efficient
implementation of periodic boundary conditions using a high-order
($p=8$) multipole local expansion. We adjust the error tolerance
parameter to limit absolute errors to 0.1\% of the rms peculiar
acceleration. As an example, a 
complete $4096^3$ particle simulation requires about
120 wall-clock hours using 12,000 CPU cores. Initial conditions were
generated using a power spectrum calculated with {\tt CLASS}~\citep{blas11}
and realized with a modified version of {\tt 2LPTIC}~\citep{crocce06}.

This particular simulation assumed WMAP 7-year cosmological 
parameters~\citep{Komatsu2011}. The box size was 1~\gmpc~on a side 
and contained $1024^3$ particles, giving a particle mass resolution 
of $7.36 \times 10^{10} \hmsol$.
All analysis in this work used a single snapshot at 
$z=0$.
For the dark matter analysis, we take successive random
subsamples of the particles to achieve tracer densities of
$10^{-2}$, $4 \times 10^{-3}$, and $3 \times 10^{-4}$ particles per 
cubic \hmpc. These samples are labeled as \emph{DM Full}, 
\emph{DM Dense}, and \emph{DM Sparse}, respectively.

We use the {\tt Rockstar} halo finder~\citep{behroozi13}, a 
six-dimensional phase-space plus
time halo finder, to identify spherical overdensity (SO) halos at 200
times the background density.  We use the default Rockstar parameters,
except for requiring strict SO masses which includes unbound particles
and particles which may exist outside of the FOF group for the halo.
We use the halo catalog both for using the halo positions as a set 
of tracers for void finding and as inputs for the HOD modeling.
We take two halo populations: one, labeled \emph{Halos Dense}, which 
uses all halos down to the minimum resolvable halo mass of 
$1.47 \times 10^{12} \hmsol$ (20 particles), and one, 
labeled \emph{Halos Sparse},
which only includes halos above $1.2 \times 10^{13}~\hmsol$. 
These samples do not have exactly the same densities as the 
HOD mocks below; rather, 
we use these two thresholds to approximate the minimum mass used 
in the HOD distribution of central galaxies, 
thereby allowing us to compare voids found 
in halos to those found in galaxy populations.

We produce galaxy catalogs from the above halo population 
using the code described in~\citet{Tinker2006} and the HOD model 
described in~\citet{Zheng2007}. 
HOD modeling assigns central and satellite galaxies to a dark matter 
halo of mass $M$ according to a parametrized distribution.
In the case of the~\citet{Zheng2007} parametrization, the 
mean number of central galaxies is given by
\begin{equation}
\left\langle N_{\rm cen}(M)\right\rangle = \frac{1}{2} \left[
1 + \erf \left(\frac{\log M - \log M_{\rm min}}{\sigma_{\log M}}\right)
\right]
\end{equation}
and the mean number of satellites is given by
\begin{equation}
\left< N_{\rm sat}(M)\right> = \left\langle N_{\rm cen}(M) \right\rangle
\left( \frac{M-M_0}{M_1'}\right)^\alpha,
\end{equation}
where $M_{\rm min}$, $\sigma_{\log M}$, $M_0$, $M_1'$, and $\alpha$ 
are free parameters that must be fitted to a given survey.
The probability distribution of central galaxies is a nearest-integer 
distribution (i.e., all halos above a given mass threshold host a central 
galaxy), and satellites follow Poisson statistics.

Using the above model, we generate two mock catalogs.
One catalog is matched to the 
number density and clustering of the 
SDSS DR9 CMASS galaxy sample~\citep{Dawson2013} using the parameters 
found by~\citet{Manera2013} ($\sigma_{\log M} = 0.596$, $M_0 = 1.2\times10^{13} \hmsol$, $M_1' = 10^{14} \hmsol$, $\alpha = 1.0127$, and $M_{\rm min}$ chosen to fit the mean number density). 
We call this sample \emph{HOD Sparse} 
since we are using it to represent a relatively low-resolution 
galaxy sample with density $3 \times 10^{-4}$ 
particles per cubic~\hmpc.
Our second catalog, named \emph{HOD Dense}, 
is matched to the SDSS DR7 main sample~\citep{Strauss2002} 
at $z < 0.1$
using one set of parameters found 
by~\citet{Zehavi2011}
($\sigma_{\log M} = 0.21$, $M_0 = 6.7\times10^{11} \hmsol$, $M_1' = 2.8 \times 10^{13} \hmsol$, $\alpha = 1.12$)
. While our simulation 
does not quite have sufficient resolution to have enough small halos
to fully capture the density of this survey, 
we will still use it to represent relatively high-resolution
galaxy samples. The density here, $4 \times 10^{-3}$ particles 
per cubic~\hmpc, is roughly 
a factor of two lower than the full DR7 density at $z < 0.1$.

\subsection{Void Finding}
We identify voids with a modified version of 
{\tt ZOBOV}~\citep{Neyrinck2008, LavauxGuilhem2011, Sutter2012a}.
{\tt ZOBOV} creates a Voronoi tessellation of the tracer particle 
population and uses the watershed transform to group Voronoi 
cells into zones and voids~\citep{Platen2007}. 
The watershed transform identifies catchment basins 
as the cores of voids and ridgelines, which separate the flow 
of water, as the boundaries of voids.
The watershed transform naturally builds a nested hierarchy of 
voids~\citep{LavauxGuilhem2011, Bos2012}, and for the purposes 
of this work ---
with the exception of Section~\ref{sec:tree} where 
we explicitly discuss the void hierarchy ---
we will only examine \emph{root} voids, which 
are voids at the base of the tree hierarchy and hence have no parents.
We also impose two density-based criteria on our void catalog. 
The first is a threshold cut within {\tt ZOBOV} itself where 
voids only include as members Voronoi cells with density 
less than $0.2$ the mean particle density.
We apply the second density criterion as a post-processing step:
we only include voids with mean central densities below $0.2$ 
the mean particle density. We measure this central density 
within a sphere with radius $R = 1/4 R_{\rm eff}$, where
\begin{equation}
  R_{\rm eff} \equiv \left( \frac{3}{4 \pi} V \right)^{1/3}.
\end {equation}
In the expression above, $V$ is the total volume of the void.
We also ignore voids with $R_{\rm eff}$ below the mean particle 
spacing of the tracer population. 

Additionally, for the analysis below we need to define a center for the 
void. For our work, we take the barycenter, or volume-weighted 
center of all the Voronoi cells in the void:
\begin{equation}
  {\bf X}_v = \frac{1}{\sum_i V_i} \sum_i {\bf x}_i V_i,
\label{eq:barycenter}
\end{equation}
where ${\bf x}_i$ and $V_i$ are the positions and Voronoi volumes of 
each tracer $i$, respectively.

Table~\ref{tab:voidsamples} summarizes the samples used in this work, 
their minimum effective void radius, 
and the number of voids identified in the simulation 
volume.

\begin{table}
\centering
\caption{Summary of sample void populations.}
\tabcolsep=0.11cm
\footnotesize
\begin{tabular}{ccccc}
  Dataset Type & Sample Name & $R_{{\rm eff}, {\rm min}}$ (\hmpc) & $N_{\rm voids}$ &\\
  \hline  \hline
DM & DM Full & 5 & 42948 & \\ 
\hline
DM & DM Dense & 7 & 21865 & \\ 
Halos & Halos Dense & 7 & 11419 & \\ 
HOD & HOD Dense & 7 & 9503 & \\ 
\hline
DM & DM Sparse & 14 & 2611 & \\ 
Halos & Halos Sparse & 14 & 2073 & \\ 
HOD & HOD Sparse & 14 & 1422 & \\ 
\hline
\end{tabular}
\label{tab:voidsamples}
\end{table}

\section{Void Hierarchies}
\label{sec:tree}

As discussed in~\citet{LavauxGuilhem2011} and~\citet{Aragon2012}, 
watershed void finders naturally 
group voids into a nested hierarchy of parents and children. 
We may use this hierarchy to understand the effects of subsampling 
and biasing before we turn to the observational consequences.
We define the 
void tree such that each void has only a single parent (or no parents at all)
and can potentially have many children. One void is considered a parent of 
another if it shares all zones of the child plus at least one more. 
Parents can then become children of even larger voids. Without any 
density thresholds, there will be a single void that encompasses the entire
simulation volume and that serves as root void for the entire population.
However, since we do apply a density threshold, we have multiple root 
voids.

\begin{figure} 
  \centering 
  {\includegraphics[type=png,ext=.png,read=.png,width=\columnwidth]{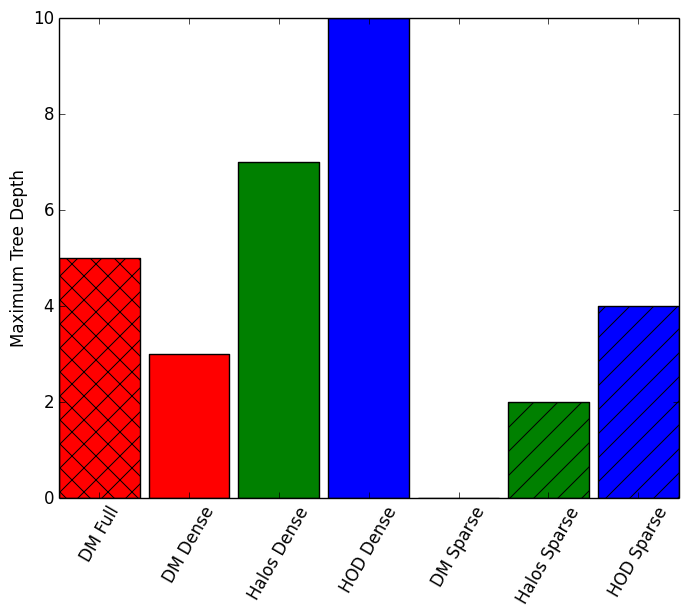}}
  \caption{Maximum tree depth in the void hierarchy for each sample.
   In addition to the axis labels, bars are colored by sample type:
   red for dark matter, green for halos, and blue for galaxies. 
   Double-hatched bars represent \emph{Full} resolution samples, 
    \emph{Dense} samples are shown with no hatching, and 
   \emph{Sparse} sample are shown with single-line hatches. 
   Subsampling of the dark matter destroys the void hierarchy, while 
   using biased tracers such as galaxies and halos reinforces the 
    hierarchy.}
\label{fig:maxtreedepth}
\end{figure}

Figure~\ref{fig:maxtreedepth} shows the maximum tree depth for each 
sample. The maximum tree depth is the length from root to tip of the 
tallest tree in the hierarchy, 
and gives a measure of the amount of substructure in the 
most complex void in the sample.
For comparison, the hierarchical tree depth of voids in a Poisson 
distribution of particles is identically zero. This follows from the 
high improbability of producing large voids in a Poisson distribution, 
and from the fact that substructures arise naturally from hierarchical 
formation.

We immediately notice the effects of lowering the sampling 
density: we completely erase any substructure information in 
the \emph{DM Sparse} sample. 
In contrast, even though the halo and galaxy populations have overall 
lower mean density, since they are \emph{biased} tracers, they 
naturally more strongly trace the substructure. 
Unsurprisingly, lower-density galaxy (and halo) populations 
have less substructure than their high-density counterparts.
Additionally, galaxy populations, which are based on their respective 
halo catalogs but have more tracers, display more substructure. 

\begin{figure} 
  \centering 
  {\includegraphics[type=png,ext=.png,read=.png,width=\columnwidth]{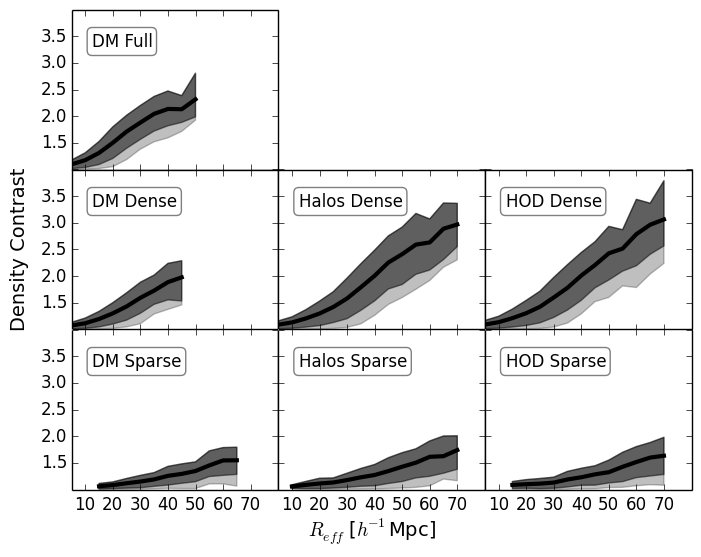}}
  \caption{Void density contrast (ratio between the mean density of particles 
           along the wall to the density of the least-dense particle) 
           versus void effective radius. 
           Solid lines are the mean of the density contrast in thin 
           bins of $R_{\rm eff}$ and dark (light) 
           grey bands are $1\sigma$ ($2\sigma$)
           ranges of the distribution in those bins.
           All samples 
           show a trend of increasing density contrast with void size.
           Halos and galaxies 
           produce very high density contrasts since many tracers
           are placed along the filaments and walls.}
\label{fig:densconradius}
\end{figure}

We can understand these results by looking at the density contrasts 
of the void populations. We define the density contrast of a void as the 
ratio between the mean density of particles along the wall to the 
density of the least-dense (or ``core'') particle.
In {\tt ZOBOV}, particles along the void wall are easy to identify:
they are adjacent to at least one non-void particle.
Figure~\ref{fig:densconradius} shows the density contrast versus 
void effective radius for each sample. 
While there is considerable scatter in the relations,
the density contrasts 
of our highest-density sample, \emph{DM Full}, clearly 
show steep dependence 
on void radius. 
We might naively expect the opposite trend, because smaller 
voids tend to have higher walls due to the \emph{void-in-cloud}
process~\citep{Sheth2004}.
However, even though the larger voids contain significant 
substructure, they tend to 
have very deep underdense subregions inside 
them, thus producing larger maximum  
contrasts.
As we lower the sampling density we find larger 
voids (which we will discuss in more detail below) but also 
much shallower voids: the maximum density contrast in the 
\emph{DM Sparse} sample is only $\sim 1/3$ the maximum contrast 
in the \emph{DM Full} sample. This behavior is entirely expected: 
as we remove particles, we ``thin out'' the walls that define 
the void boundaries and thus reduce the overall density at the 
edges.

However, both the halo and HOD samples,
 even though they contain fewer tracers than the 
high-resolution dark matter cases, have comparable --- and even higher --- 
density contrasts than the low-resolution dark matter samples.
We expect this because of biasing: we generally make the walls 
thinner by switching to halos and galaxies, 
leading to larger voids, but the walls that survive are 
more densely concentrated 
with tracers, leading to high density contrasts.
The relationship between density contrast and radius is nearly 
identical between voids in galaxies and halos, since they are 
both already biased tracers.

We may also consider
the position of the void in the 
tree hierarchy. All 
root voids have a broad range of density contrasts, since these 
include isolated small voids with low density contrasts and large
voids that serve as parents of sub-voids.
On the other hand, for all samples, the most deeply-nested voids 
exclusively have low density contrasts. 
Thus we clearly see that as we lower the sampling density, 
we puncture lower-density walls, allowing the watershed algorithm to 
join adjacent basins, and more 
preferentially erase smaller, deeply-nested voids.
Voids that happen to have high-density walls tend to survive. 
The HOD and halo samples are able to recover substructure 
by selectively placing tracers along the walls and filaments that
separate voids. 


\section{Void Statistics}
\label{sec:observable}

\subsection{Number functions}
Perhaps the most fundamental void statistic is the void number function: 
the number of voids of a given size per unit volume. 
Since the growth of voids is intimately tied to the growth 
of structure, which is itself controlled by the nature of dark energy and the 
amount of dark matter in the universe, the void number function 
is sensitive to cosmological parameters~\citep[e.g.,][]{Jennings2013}.
Since different cosmologies affect the number function in different 
ways (e.g., by suppressing the formation of larger voids 
and thereby tilting the distribution), we must disentangle the effects of 
sparsity and biasing, which can modify the number function in similar ways.

\begin{figure} 
  \centering 
  {\includegraphics[type=png,ext=.png,read=.png,width=\columnwidth]{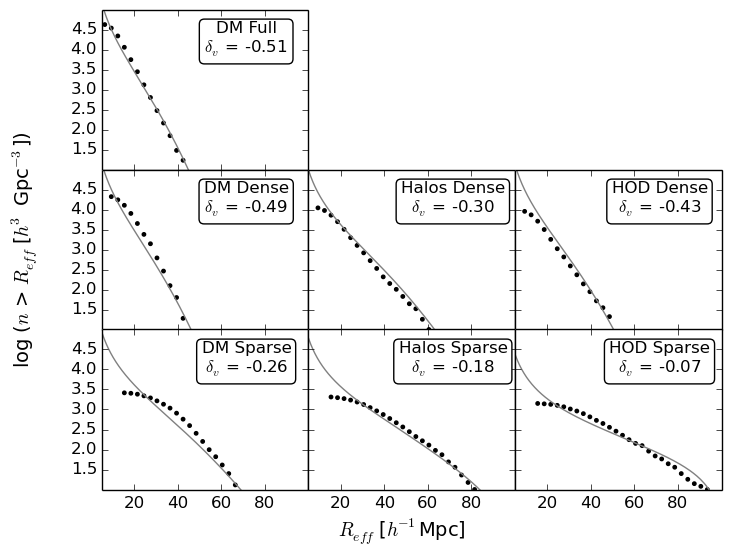}}
  \caption{Cumulative void number function. 
           The dotted lines are measured 
           number functions in each indicated sample, and the thin 
           grey lines 
           are theoretical predictions of the SVdW function with the 
           $\delta_v$ parameter adjusted to the indicated value. 
           The best-fit parameter value is shown in each sub-plot.
           }
\label{fig:numberfunc}
\end{figure}

Figure~\ref{fig:numberfunc} shows our cumulative void number function
for each sample.
As previous authors~\citep{Colberg2005, Jennings2013, Watson2013} 
have noted, voids 
in lower-density samples tend to be larger, and there tend to be 
fewer small voids.
This is due to two causes. First, the walls and filaments 
are thinner in low-density samples, leading to the same 
population of voids as in high-density samples, but the individual voids 
are just larger. Secondly, if a lower-density wall separating 
two voids loses too many 
particles due to subsampling, the watershed method will 
merge the two basins without first identifying them as 
separate voids. In~\citet{Sutter2013b}, 
we examine the relative importance of these two effects for different 
galaxy sampling densities. For this work, we ignore the underlying 
causes and focus on the observational consequences.

We see that the void number functions from halo catalogs closely follow
that of the mock galaxy catalogs, as we expected from the discussion 
in the previous section. 
The difference between the \emph{DM Sparse} and \emph{HOD Sparse}
is due to the interaction between biasing and our central density cut. At fixed tracer density a more biased tracer will lead to a sharper density contrast for a given size of void. 
Biasing concentrates tracers along the walls enhancing the density contrast, see Figure~\ref{fig:densconradius} and the associated discussion. Therefore large voids 
found in biased tracers will tend to survive a central density cut in spite of the voids being partially filled with dark matter.
The corresponding shallow underdensities in the dark matter sample have more substructure, and will therefore not survive the density cut. In the high-resolution cases the substructure remains even in the biased tracers, rendering this effect much less pronounced.

To provide theoretical number functions, we turn to the analysis of Sheth \& van de Weygaert (2004) (hereafter SVdW), who employ the excursion-set formalism to develop a void number function of the form
\begin{equation}
n(M) = \frac{\bar{\rho}}{M^2}\nu f(\nu)\frac{\mathrm{d}\ln\nu}{\mathrm{d}\ln M}\;,
\label{eq:numfunc}
\end{equation}
where $\bar{\rho}$ is the background matter density, $M$ the void mass, $\nu=\delta_v^2/\sigma^2(M)$ with $\delta_v$ being the critical underdensity for void formation and $\sigma^2(M)$ the variance of the density field on a scale $R\simeq(3M/4\pi\bar{\rho})^{1/3}$, and
\begin{equation}
\nu f(\nu) = \sqrt{\frac{\nu}{2\pi}}\exp\left(-\frac{\nu}{2}\right)\exp\left(-\frac{|\delta_v|}{\delta_c}\frac{\mathcal{D}^2}{4\nu}-2\frac{\mathcal{D}^4}{\nu^2}\right)\;. 
\end{equation}
Here, $\mathcal{D}\equiv|\delta_v|/(\delta_c+|\delta_v|)$ is the
 so-called \emph{void-and-cloud} parameter, 
which determines the relative importance of halo and void formation. 
The value for the critical overdensity of collapse $\delta_c\simeq1.686$ 
is tightly constrained by the spherical collapse model, 
but Furlanetto \& Piran (2006) pointed out that the value 
of $\delta_v$ must be adjusted to account for the sampling density of the tracers used to identify voids.

While there have been attempts to improve this basic 
relation~\citep{Paranjape2012, Jennings2013, Achitouv2013}, 
they still require parameter 
adjustments to translate between void populations in dark matter and 
galaxies. Until the development of a robust dynamics-based 
prediction for the void number function that automatically accounts 
for sparsity and biasing, we should take these 
models as a basis for phenomenologically fitting to void populations 
in dark matter and mock galaxy catalogs.

In that spirit, we employ the fixed radial rescaling of $1.7$ described in
~\citet{Jennings2013} and adjust the $\delta_v$ parameter to attempt 
to fit our measured void number functions. 
Attempts to rescale the radii beyond the fixed valued of $1.7$ 
did not produce better fits.
We show these best-fit 
values in Figure~\ref{fig:numberfunc}.
While the simple functional form of Eq.(\ref{eq:numfunc})
 matches the rough order of magnitude and approximate shape
for each sample, it misses the detailed shape, and thus we do not 
evaluate a goodness-of-fit. 

Although we are unable to match the detailed shape of the number 
function with the SVdW functions, we do match the gross properties 
and overall number counts with surprising ease: a single parameter 
adjustment allows us to approximate the number functions in both high- 
and low-resolution samples of all tracer types, although it is 
more difficult to account for the effect of biasing on large voids.
We see that the void number function is more dependent on the 
density of tracers rather than the type of tracers, and that 
we can divide our samples into low-resolution and high-resolution 
groups. 

The values of $\delta_{v}$ shown  in Figure \ref{fig:numberfunc} 
 stand in contrast to the expected $\delta_v = -2.8$
based on shell-crossing criteria for spherical underdensities
~\citep{Sheth2004,Biswas2010}. Two factors contribute to this difference: our void finder discovers the full non-spherical 
shape of the underdensity, so the spherical approximation breaks down, 
and the assumption of complete shell-crossing across the entire 
void surface is too restrictive in general~\citep{Falck2012, Neyrinck2013}. 

\subsection{Ellipticity Distributions}

There has been considerable interest recently in the dependence of 
void shapes on the nature of dark 
energy~\citep[e.g.,][]{Lee2006, Park2007, Biswas2010, Bos2012}. 
The shape distribution of voids is the inverse of the growth of structure: 
as matter collapses to form galaxies, groups, and clusters, the 
shapes of the evacuated regions will necessarily change. Since 
measurements of voids are largely unaffected by systematics 
due to baryonic physics, the redshift evolution of the void 
shape distribution potentially serves as a powerful 
tracer of dark energy. In addition, the mean stretch of 
voids along the line of sight may be used 
for an application of the  Alcock-Paczynski 
test~\citep{Alcock1979, Ryden1995, LavauxGuilhem2011, Sutter2012b}.

We will simplify the discussion of void shapes by focusing 
on the ellipticity. To compute the ellipticity, for a given
set of tracers within a void we first construct the 
inertia tensor:
\begin{eqnarray}
  M_{xx} & = &\sum_{i=1}^{N_p} (y_i^2 + z_i^2) \\ 
  M_{xy} & = & - \sum_{i=1}^{N_p} x_i y_i, \nonumber
\end{eqnarray}
where $N_p$ is the number of particles in the void, and 
$x_i$, $y_i$, and $z_i$ are coordinates of the particle $i$ 
relative to the void barycenter.
The other components of the tensor are obtained by 
cyclic permutations.
Given the inertia tensor, we compute the eigenvalues and form
the ellipticity:
\begin{equation}
  \epsilon = 1- \left( \frac{J_1}{J_3}\right)^{1/4},
\label{eq:ellip}
\end{equation}
where $J_1$ and $J_3$ are the smallest and largest eigenvalues, 
respectively. 

\begin{figure}
  \centering 
  {\includegraphics[type=png,ext=.png,read=.png,width=\columnwidth]{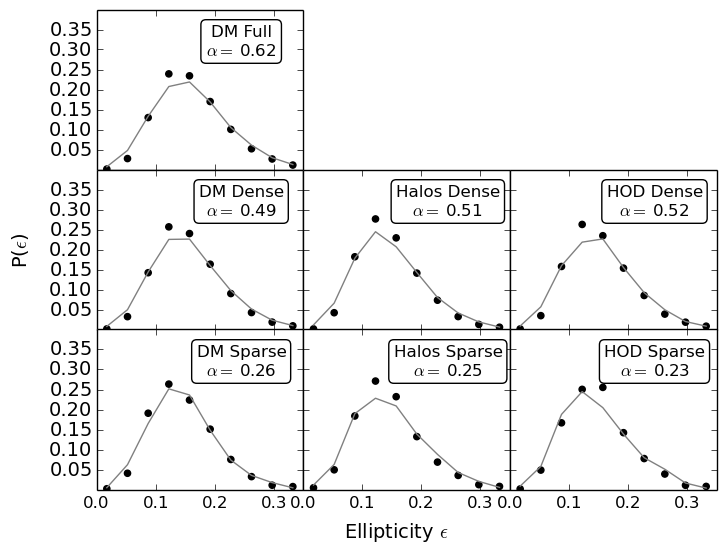}}
  \caption{Measured ellipticity distributions (points) for each sample 
           and best-fit theoretical distributions (thin grey lines). 
           To produce the theoretical distributions we use
           {\tt DIVA} with the rescaling parameter shown in each subplot.
           This rescaling 
           parameter $\alpha$ (Eq.~\ref{eq:r0}) sets the scaling from 
           Eulerian to Lagrangian scales}. 
\label{fig:ellip}
\end{figure}

Figure~\ref{fig:ellip} shows the ellipticity distribution 
for each sample. We 
contrasted these ellipticities to the ellipticities of 
voids founds in a random Poisson distribution of particles with 
density $10^{-2}$ tracers per cubic~\hmpc.
Voids in all samples are more elliptical and have broader distributions 
than voids in random particles. For each type of tracer, 
the amount of ellipticity 
of the voids correlates with the density of the underlying tracer 
population: lower density tracers produce more spherical 
voids. This is mainly due to the loss of the smallest voids, 
which tend to have higher ellipticities~\citep{Sutter2013c} 
than medium-scale voids, which survive at lower sampling density.

Comparing samples at fixed density 
against each other, we see that 
halos produce more spherical voids while galaxies recover 
similar ellipticities as the original dark matter samples, 
although for the  \emph{Sparse} samples the ellipticities 
are largely indistinguishable.
The halo populations produce more spherical voids because 
of the lack of tracers for the smallest void populations.
However, the presence of satellites in the HOD galaxy populations
generally adds more tracers along the void boundaries giving 
better shape measurements, and hence higher ellipticities. 
This same phenomenon --- that of increased hierarchical structure 
in the \emph{HOD} population --- gives rise to the 
more substantial tree structure in Figure~\ref{fig:maxtreedepth}.
Thus results such as~\citet{Bos2012}, 
which use voids in halo populations to assess the power of 
void ellipticities in galaxy surveys to constrain exotic dark energy models, 
may be too pessimistic.

To compare to theoretical expectations we use the Lagrangian model of
Lavaux \& Wandelt (2010) developed in the context of 
{\tt DIVA} (DynamiCal Void Analysis), 
a Lagrangian void finder and characterizer.
{\tt DIVA} is based on the capability to compute the Lagrangian comoving
coordinates of any particles or galaxies from solely its Eulerian
position. This capability is provided by the Monge-Ampere-Kantorovitch
(MAK) reconstruction~\citep{Brenier2003,Lavaux2008}. From the
mass tracer distribution at a given redshift $z$, MAK finds the proper
change of coordinates to map the position of the tracers at high
redshift such that the density becomes totally homogeneous. Through this
transformation, we may model very accurately the distribution and the
evolution of ellipticity with redshift.

Lavaux \& Wandelt (2010) showed that to first order the inertial tensor
estimate of the ellipticity and its counterpart estimated from dynamics
are equivalent, though with large scatter. We compare the analytical
model of the evolution of the ellipticity to our measured ones obtained
from the inertial tensor method, neglecting the impact of the scatter.
To do so, we need to select a proper smoothing scale for the Lagrangian
calculation in {\tt DIVA}, and the link between this smoothing scale and the
Eulerian radius that we measure in simulations and data is not clear.
For simplicity, we will assume a simple scaling in which 
for a void of given radius $R_{\rm eff}$ we choose a smoothing 
scale
\begin{equation}
  R_0 = \alpha R_{\rm eff},
\label{eq:r0}
\end{equation}
where $\alpha$ is a free parameter.
As discussed in~\citet{Lavaux2010}, we may identify the \emph{Eulerian} 
ellipticity $\epsilon$ with the \emph{Lagrangian} eigenvalues $\lambda_i$ of the
gravitational shear tensor via
\begin{equation}
  \epsilon \cong 1 - \left( \frac{1+\lambda_1}{1+\lambda_3}\right)^{1/2},
\label{eq:diva_ellip}
\end{equation}
where $\lambda_1$  is the smallest eigenvalue and $\lambda_3$ is the 
largest.
For each individual void, we choose the smoothing scale with Eq.(\ref{eq:r0}) 
and
compute the distribution of ellipticities (Eq.~\ref{eq:diva_ellip}).
We then draw an ellipticity from a Gaussian distribution with the same 
mean and variance. This is simply for numerical convenience but
approximates well the actual distribution. 
We take this randomly-chosen ellipticity as the predicted ellipticity 
of that void.
Finally, we build a distribution of the predicted 
ellipticities for comparison to the distribution of ellipticities 
as measured in our simulation with Eq.(\ref{eq:ellip}).

Figure~\ref{fig:ellip} shows the measured ellipticity distributions 
for each sample as well as the theoretical distributions with 
best-fit values of $\alpha$. 
We found that for higher-resolution samples 
(\emph{DM Full} and \emph{HOD Dense}), a choice of $\alpha=0.5-0.6$ 
produced 
theoretical distributions in excellent agreement with the measured 
distributions: not only do we match the mean ellipticities, 
but also we broadly recover the shapes of the distributions. 
For sparse samples, a value of $\alpha \sim 0.25$ was more appropriate.
This change in the parameter $\alpha$ to accommodate low-resolution 
samples is analogous to the need to change $\delta_v$ for number functions. 
In this case, lower values of $\alpha$ indicate that the dynamical 
cores of voids in the \emph{Sparse} samples are smaller than in the 
higher-resolution samples, in line with the preceding number function 
analysis. 

\subsection{Radial Density Profiles}

The radial density profiles of voids are exceptionally sensitive to 
modified gravity and fifth forces, since the absence of matter 
allows exotic forces to remain 
unscreened~\citep{Clampitt2013,Spolyar2013}.
However, since the choice of mass tracer can also affect density profiles
~\citep{Benson2003,Colberg2005,Padilla2005}, we must first understand 
the impacts of sparsity and biasing before using voids as probes of fundamental physics. 
If we can provide a translation from theoretical predictions in 
dark matter to observational radial profiles we can forecast if
modifications due to exotic forces will persist when using 
galaxies to define voids.

\begin{figure*} 
  \centering 
  {\includegraphics[type=png,ext=.png,read=.png,width=0.48\textwidth]{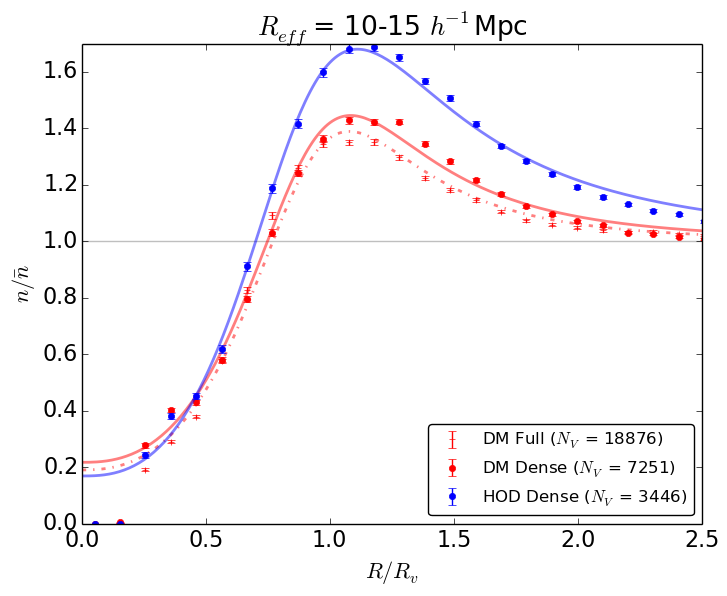}}
  {\includegraphics[type=png,ext=.png,read=.png,width=0.48\textwidth]{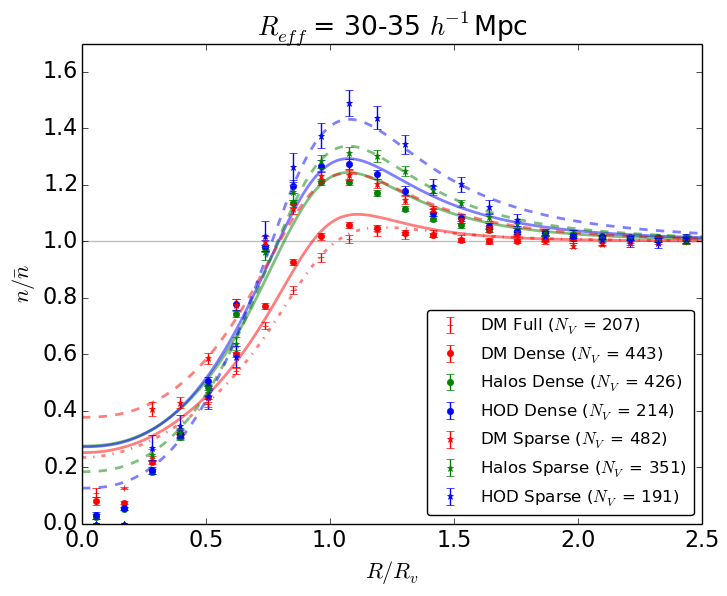}}
  {\includegraphics[type=png,ext=.png,read=.png,width=0.48\textwidth]{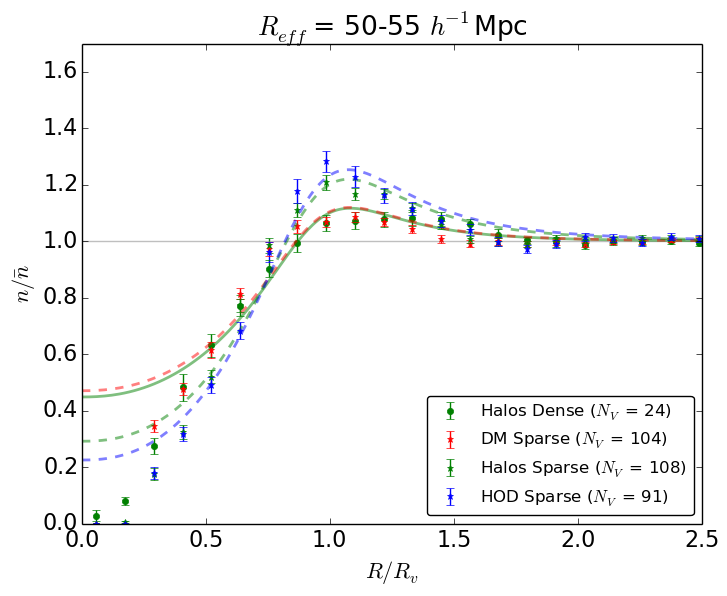}}
  {\includegraphics[type=png,ext=.png,read=.png,width=0.48\textwidth]{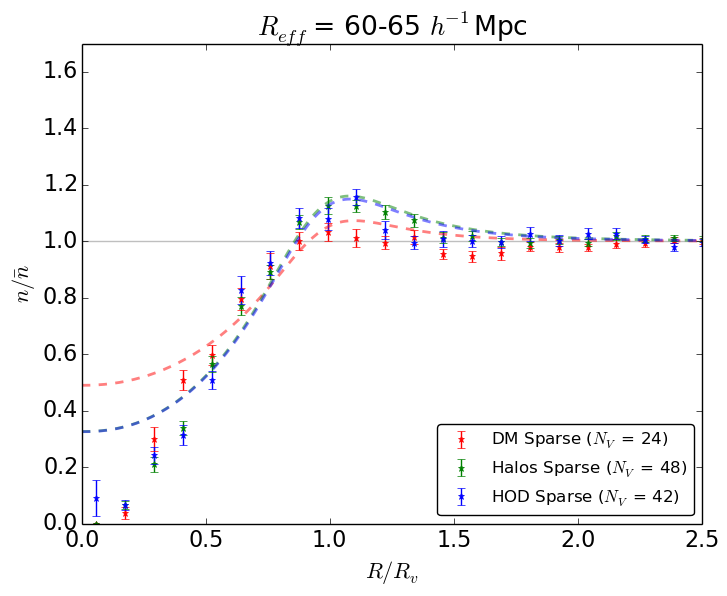}}
  \caption{One-dimensional radial density profiles of stacked voids 
           (points with error bars) and best-fit curves (thin lines) 
           using the profile
           discussed in Eq.(\ref{eq:profile}).
           For the fit we ignore the first three bins, since these are
           artificially influenced by numerical effects in the simulation and 
           the application of the central density cut.
           Each profile is normalized to the mean number density $\bar{n}$
           of that sample and $R_v$ corresponds to the median
           void size in the stack.
           Solid colored lines are from high-density samples, 
           and dotted colored lines are from low-density samples.
           Each tracer type is given a unique color and 
           symbol as indicated in the legend.
           The best-fit values are plotted in Figure~\ref{fig:profile_fits}.
           }
\label{fig:1d_profile}
\end{figure*}

Figure~\ref{fig:1d_profile} shows one-dimensional radial profiles 
for all samples in a few selected radius ranges. 
To compute the profiles, we take all voids in a 
sample of a
given size range (e.g., $30-35$ \hmpc), align all their barycenters, 
and measure the density in thin spherical shells. We normalize each 
density profile to the mean number density of the sample and show 
all profiles as a function of relative radius, $R/R_v$, where $R_v$ 
is the median void size in the stack.

Many authors have discussed and presented measured radial density profiles 
in data and simulations, and there appears to be a generally 
universal shape to the profile: a central underdense floor, 
a steep wall, a slightly-overdense compensation, and a declining 
density that asymptotes to the mean 
density~\citep[e.g.,][]{Benson2003, Padilla2005, Ceccarelli2006, 
LavauxGuilhem2011, Sutter2012a}. 
The \emph{internal} density profiles of voids are nearly identical across
all samples, with lower-density samples producing slightly steeper slopes, 
lending further evidence for the
 existence of this universal void profile 
and that sparsity and galaxy bias do not impact this profile much.

However, the \emph{external} void profile shows significant 
differences among the samples.
The height of the compensation bump depends on the surrounding 
medium and is generally dependent on void size: small 
voids are typically found in large overdense regions 
(the \emph{void-in-cloud} process) and large voids 
are typically found in underdense 
regions (the \emph{void-in-void} process). 
This separation of void types was first discussed 
in~\cite{Sheth2004} and measured in SDSS voids 
by~\citet{Ceccarelli2013}.

For a stack of voids of a given size range, low-density samples of all types appear 
more overcompensated and high-density samples appear 
more undercompensated with shallower slopes and lower 
compensation regions.
Comparing samples at a given density, the slopes of 
the profiles are very similar 
but with different compensation heights, due to the biasing 
of the galaxies and halos relative to the dark matter.
This may seem counterintuitive given our measurements of the 
density contrasts in Figure~\ref{fig:densconradius}. However, that 
measurement was a comparison of the density of the core particle 
(not a spherically-averaged volume as in these plots) to the 
density on the wall of the void, which is just inside the 
compensation region. 

As  with the number functions, we see that 
halos offer a good proxy for the galaxy population.
Interestingly, \emph{all} halo and galaxy populations, regardless of 
sampling density, produce very similar radial profiles. 
This is likely due to the fact that the radial profile 
is more sensitive to the biasing of the tracers than to the 
density. Indeed, in this context we can understand biasing as just 
another form of subsampling: biased but high-density populations 
(e.g., \emph{HOD Dense}) act very similarly to unbiased 
but low-density populations (e.g., \emph{DM Sparse}).
This is in contrast with the effect on number functions: there, sparsity 
was more important than biasing. However, the number functions 
probe the interior contents of voids, and this discussion has focused 
mainly on the compensation, which is in the \emph{surrounding} medium.

\begin{figure} 
  \centering 
  {\includegraphics[type=png,ext=.png,read=.png,width=\columnwidth]{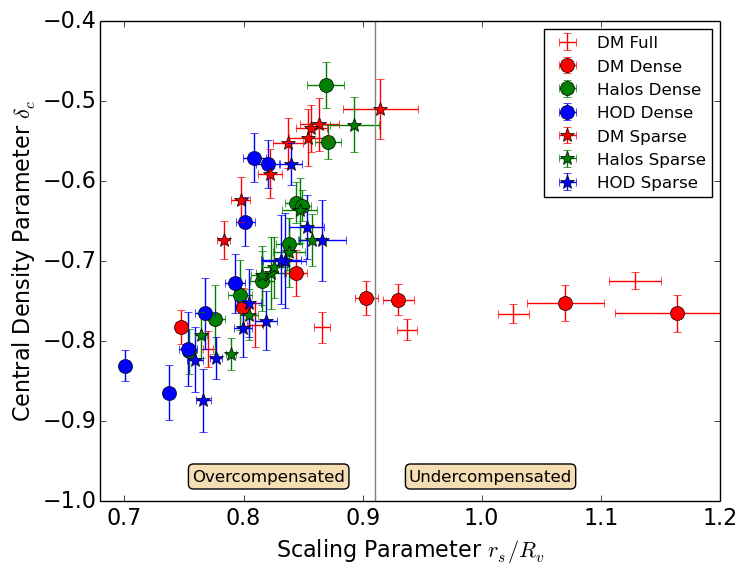}}
  \caption{Best-fit values and $1 \sigma$ uncertainties 
           for all void stacks in all samples. 
           Tracer types are distinguished by colors and densities
           by symbol, as noted in the legend. The thin grey line depicts
           the compensation scale.
           }
\label{fig:profile_fits}
\end{figure}

There is very little theoretical 
development into \emph{predictions} for the shapes of profiles. 
Given the lack of theoretical motivation, authors generally 
attempt fits to the radial density profile using an empirical formula 
that attempts to reproduce the profile shapes. 
We use the recent profile described in~\citet{Hamaus2014}, which 
includes a functional form spanning both the interior void slope and 
the compensation region:
\begin{equation}
\frac{n}{\bar{n}} (r) = \delta_c \frac{1- (r/r_s)^{\alpha(r_s)}}
                             {1 + (r/R_v)^{\beta(r_s)}} + 1,
  \label{eq:profile}
\end{equation}
where~\citet{Hamaus2014} found that 
\begin{eqnarray}
  \alpha(r_s) & \simeq & -2.0 (r_s/R_v) + 4.0 \\
\beta(r_s) & \simeq & \left\{ \begin{array}{rl}
17.5 (r_s/R_v) - 6.5 &\mbox{ if $r_s/R_v < 0.91$} \\
-9.8 (r_s/R_v) + 18.4 &\mbox{ if $r_s/R_v > 0.91$}.
\end{array} \right.
  \label{eq:parmfits}
\end{eqnarray}
There are two free parameters to this model: $r_s$, the radius 
at which the profile reached mean density, and $\delta_c$, 
the underdensity in the central core. Figure~\ref{fig:profile_fits} shows
are best-fit values of $\delta_c$ and $r_s$ for all void stacks 
in all samples.
This two-parameter model describes nearly all voids very well, although 
we see that it has difficulty reproducing the exact height of the 
compensation bump for the very largest voids.
This is due to the fact that Eq.(\ref{eq:parmfits}) is tuned to voids 
in high-resolution dark matter, and not necessarily appropriate for 
these cases. Despite this, however, the fits agree remarkably well.

We can understand these differences in terms of the void size at 
which the void bias changes sign~\citep{Hamaus2013}: the 
compensation scale. This 
change in the bias represents a switch from generally overcompensated 
to generally undercompensated voids. 
Voids in higher-density 
tracers have a smaller compensation scale than voids in low-density 
tracers. Thus, if we choose a stack of voids with radii well below this 
compensation scale for all tracers, for example $10-15$~\hmpc, 
then the profiles will look very similar. Similarly, if we stack very large 
voids we only see undercompensated voids and again the 
profiles are similar. For intermediate scales, high-density 
tracers such as \emph{DM Full} will have switched from 
over- to undercompensation, while low-density tracers will 
remain overcompensated. This manifests in the different slopes 
and compensation heights in the profiles from $\sim 20$ to $\sim 50$~\hmpc.
Note that this is a continuous progression from over- to 
under-compensation, so generally as we increase the stack size 
we see gradually reduced compensations.

\begin{figure*} 
  \centering 
  {\includegraphics[type=png,ext=.png,read=.png,width=0.48\textwidth]{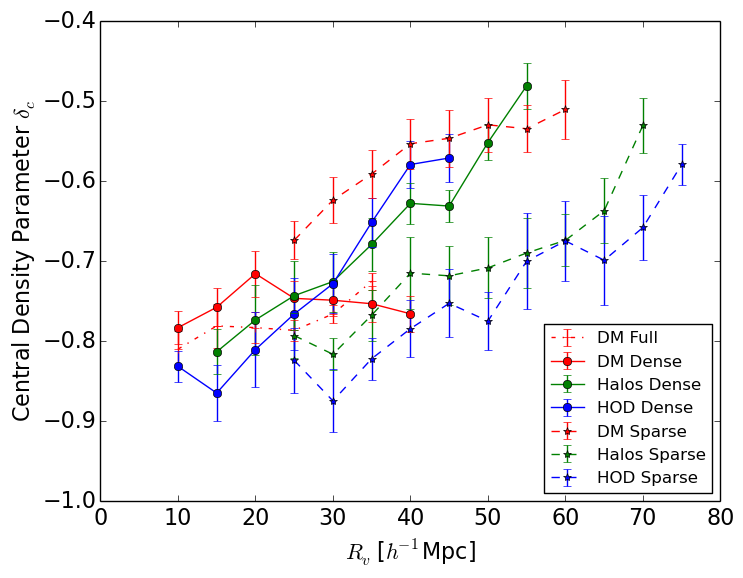}}
  {\includegraphics[type=png,ext=.png,read=.png,width=0.48\textwidth]{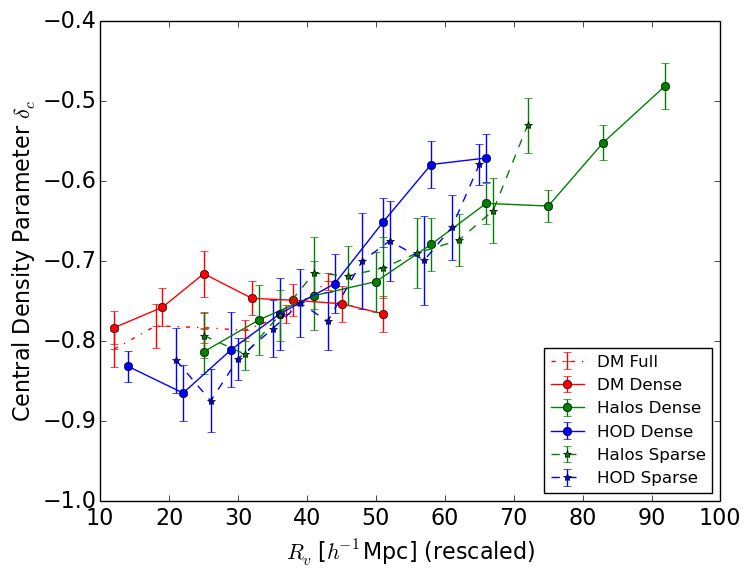}}
  {\includegraphics[type=png,ext=.png,read=.png,width=0.48\textwidth]{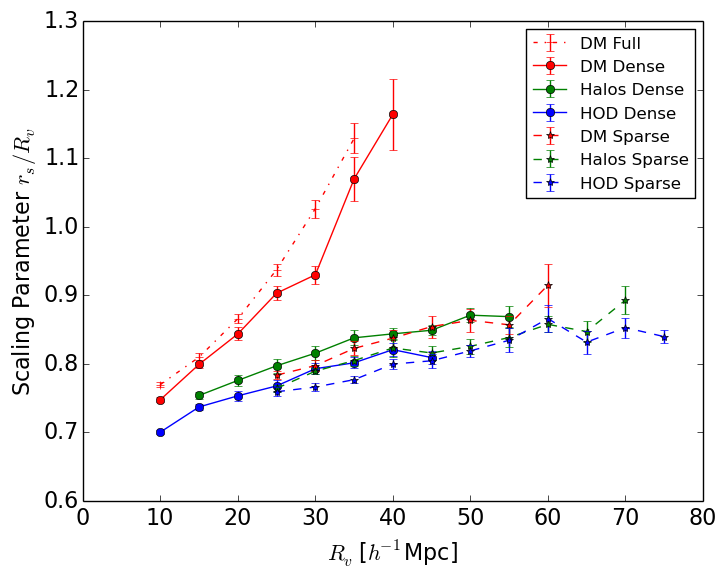}}
  {\includegraphics[type=png,ext=.png,read=.png,width=0.48\textwidth]{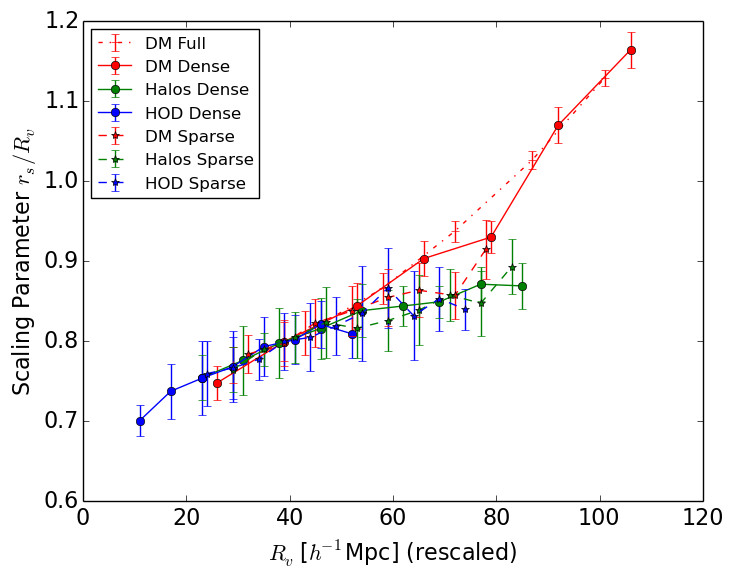}}
  \caption{Universal rescaling of voids. We show the fitting parameter 
           $\delta_c$ versus void radius $R_v$ in the top row, and the 
           bottom row shows $r_s$ versus void radius $R_v$ for each 
           sample. The left-hand column is the non-rescaled 
           fitting parameters. On the right, we pick a fiducial 
           point in the plane and rescale all voids in a given sample 
           by a single value so that the parameter-size curve passes through
           that point. After rescaling, the curves cluster around a universal relationship between 
void size and the fitting parameters 
           that is independent of tracer density and bias.
           }
\label{fig:dcvsr}
\end{figure*}

Thus, we can in principle rescale any void from any tracer population 
onto any other void from any other tracer population, and 
Figure~\ref{fig:dcvsr} demonstrates such rescaling. In this figure, 
we plot each fitting parameter separately as a function of void 
size for each sample. By choosing a single fiducial parameter value, we 
rescale all void radii in a sample with only a single scale factor. 
In other words, we can pick any parameter-radius relationship from any 
sample and shift all of the other relations on top of it by 
simply rescaling each of their void radii by a single number.
When this is performed 
for each sample, the curves cluster around a universal relationship between 
void size and the fitting parameters that is independent of 
sample density and bias. 

\section{Conclusions}
\label{sec:conclusions}

By using direct subsampling, halo finding, and HOD modeling 
on a single high-resolution, large-volume $N$-body simulation, 
we have carefully investigated the impacts of tracer density 
and biasing on void statistics.
We have used the void tree hierarchy to understand the impacts 
on number functions, ellipticities, and density profiles.
We find that sampling density is much more important for measured void 
properties than galaxy bias. As we lower the tracer density, 
voids become larger, small voids disappear, they become 
more spherical, and their profiles become slightly steeper.
The biasing due to galaxy tracers recovers some of the loss 
of information due to low densities: the tree hierarchy 
is restored and void walls become 
steeper, leading to higher ellipticities.
This kind of hierarchical tree analysis can be directly applied 
to voids in observations, since it is a natural feature of the 
watershed algorithm and already implemented in our 
approach described in~\citet{Sutter2012a}. However, the limited 
survey volumes available in the SDSS prevent a detailed examination
at this time.

We have used theoretical and empirical formulas to 
model the statistical void properties as a function of tracer 
type and density: 

\emph{Number functions -} The SVdW number function approximates well 
the behavior of measured number functions with simple adjustments to the 
``void parameter'' $\delta_v$, although further work is needed 
to model specific shapes. These adjustments make sense since the 
$\delta_v$ parameter relates the size of a void to the 
central underdensity \emph{in the dark matter}.
Since number functions of voids in low-density samples are best fit 
by lower values of $\delta_v$, this tells us that the voids 
discovered in low-density surveys, while still underdense, 
represent very shallow, wide density perturbations, as seen 
in the HOD modeling analysis of~\citet{Sutter2013b} and 
the \emph{ab initio} simulations of~\citet{Ricci2014}.

\emph{Ellipticity distributions -} 
Here semi-analytic calculations are able to describe with very high fidelity 
the statistics of void ellipticities. 
One can take a mock void population (either produced from simulations 
or from a theoretical distribution), rescale appropriately, and 
estimate the resulting ellipticity distribution with methods such 
as {\tt DIVA}. The rescaling parameter $\alpha$ must be 
adjusted to account for the 
sparsity of the survey: $\alpha \sim 0.5$ is appropriate for high-resolution 
surveys while $\alpha \sim 0.25$ is necessary for low-resolution surveys.
Once this single choice of parameter is made, there is remarkably 
good agreement in both the mean and the shape of the distributions.
This change in the required value of $\alpha$ is consistent 
with the change in $\delta_v$, since $\alpha$ connects the Eulerian 
size of a voids to its Lagrangian size, and sparser samples 
will map out larger Eulerian volumes around the same core
underdensity.

\emph{Radial profiles -} While there is limited theoretical motivation 
for any particular void profile at all scales 
(although, as discussed in~\citealt{Papai2011}, 
for large enough scales the BBKS formalism provides 
an expected profile), 
the function of~\citet{Hamaus2014} describes all voids in all samples 
to a remarkable degree of accuracy. Using this fit we have shown that voids 
obey a universal and self-similar relationship between central 
underdensity, scaling radius, and size. Thus voids in one sample 
or survey can be immediately rescaled to match voids in another 
sample or survey, even from high-resolution $N$-body simulations 
to a sparse galaxy survey. 

We have also judged the ability of halos to approximate the 
galaxy distribution, a common approach, since HOD and semi-analytic 
modeling are subtle and computationally expensive.
We found that for number functions and 
radial profiles, this is a good approximation, but \emph{only when 
an appropriate minimum mass threshold is chosen}, specifically 
by matching the abundances of central galaxies.
This is similar to the conclusions of~\citet{Padilla2005}. 
Simulations have a 
variety of volume and resolutions, and the minimum halo mass in a 
given simulation may not correspond to any particular survey. 
Care must be taken to make predictions for specific surveys.
Also, we found that voids in halo distributions are more 
spherical than in galaxy distributions: for ellipticity 
predictions, either a constant shift to account for galaxies 
or an approach like {\tt DIVA} must be used.

This work is only a first step: we have looked at sparsity and biasing, 
but ignored redshift space distortions and effects of survey masks. 
In the case of radial profiles, we may use techniques 
such as those described in~\citet{Pisani2013} to translate from 
redshift to real space and make contact with our results. 
The early work of~\citet{Ryden1996}  has looked at void shapes in 
real- versus redshift-space, but a more comprehensive study is 
needed. Our companion work~\citep{Sutter2013c} includes a
preliminary analysis of the impact of survey masks.

We have made all populations (dark matter, halos, and mock galaxies) 
 used in this work as well as the resulting void information 
publicly 
available alongside our catalog of voids in the SDSS DR7
and DR9 at {\tt http:www.cosmicvoids.net}. The formatting of the 
catalog follows that described in the Appendix 
of~\citet{Sutter2012a}. The catalog contains a {\tt REAMDE} 
with more detailed information.

While our recommendations obviously depend on specific survey 
details, we have provided broad guidelines for judging the 
feasibility of a particular void statistic to inform us about 
cosmology. By quantifying the effects of sparsity and biasing, 
we can translate between results in theory, high-resolution 
dark matter, and current and future galaxy surveys.
Also, our discovery of a simple rescaling of void density profiles 
suggests that we may study voids in dark matter and trivially 
make predictions for voids in galaxy surveys, regardless of 
sampling density and bias.
This allows future work to disentangle these effects from 
cosmological signals,
opening the 
way for more effective and straightforward void cosmology predictions.

\section*{Acknowledgments}

PMS, NH, and BDW acknowledge
support from NSF Grant NSF AST 09-08693 ARRA. BDW
acknowledges funding from an ANR Chaire d'Excellence (ANR-10-CEXC-004-01),
the UPMC Chaire Internationale in Theoretical Cosmology, and NSF grants AST-0908
902 and AST-0708849.
GL acknowledges support from CITA National Fellowship and financial
support from the Government of Canada Post-Doctoral Research Fellowship.
Research at Perimeter Institute is supported by the Government of Canada
through Industry Canada
 and by the Province of Ontario through the Ministry of Research and
Innovation.
DW acknowledges support from NSF Grant AST-1009505.
This work made in the ILP LABEX (under reference ANR-10-LABX-63) was supported by French state funds managed by the ANR within the Investissements d'Avenir programme under reference ANR-11-IDEX-0004-02.

\footnotesize{
  \bibliographystyle{mn2e}
  \bibliography{voidmocks}
}

\end{document}